# Direct observation of silver nanoparticle-ubiquitin corona formation

*Feng Ding[1,2,*], Slaven Radic[1], Poonam Choudhary[1], Ran Chen[1], Jared M Brown[3] and Pu Chun Ke[1,*]*

[1]Department of Physics and Astronomy, Clemson University, [2]Deapartment of Biochemistry and Biophysics, School of Medicine, University of North Carolina at Chapel Hill, [3]Department of Pharmacology and Toxicology, Brody School of Medicine, East Carolina University

**ABSTRACT**

Upon entering physiological environments, nanoparticles readily assume the form of a nanoparticle-protein corona that dictates their biological identity. Understanding the structure and dynamics of nanoparticle-protein corona is essential for predicting the fate, transport, and toxicity of nanomaterials in living systems and for enabling the vast applications of nanomedicine. We combined multiscale molecular dynamics simulations and complementary experiments to characterize the silver nanoparticle-ubiquitin corona formation. Specifically, ubiquitins competed with citrates for the nanoparticle surface and bound to the particle in a specific manner. Under a high protein/nanoparticle stoichiometry, ubiquitions formed a multi-layer corona on the particle surface. The binding exhibited an unusual stretched-exponential behavior, suggesting a rich kinetics originated from protein-protein, protein-citrate, and protein-nanoparticle interactions. Furthermore, the binding destabilized the α-helices while increasing the β-sheets of the proteins. Our results revealed the structural and dynamic complexities of nanoparticle-protein corona formation and shed light on the origin of nanotoxicity.

Nanomaterials have been increasingly applied in consumer products due to their unique physical and chemical properties. The increasing application of nanomaterials in daily life inevitably leads to their accumulation in the environment[1] and subsequent entry into biological systems, causing bio-safety concerns related to nanotechnology[2]. Nanoparticles have also been found useful in disease diagnostics, drug and gene delivery, and therapeutics[3-5]. Therefore, the safety issue of nanotechnology is pressing, and the study of nanotoxicology has attracted much research interest recently[6]. The benefits of understanding the interactions between nanoparticles and biological systems extend from fundamental physical sciences to nanomedicine, nanotoxicology, nanoecotoxicology, consumer usages, and the public's perception of nanotechnology.

Upon entering biological systems such as the bloodstream, a nanoparticle forms molecular complexes with encountered proteins, termed as the protein corona[7]. Protein corona shields the surface of the exogenous nanoparticle and subsequently determines the biological properties of the nanoparticle core. On the other hand, interactions with nanoparticles can also alter the structure, dynamics, and function of the bound proteins, which could further impact recognition of the proteins by membrane receptors and the immune system. Previous experimental studies have provided much insight, such as the existence and size of the protein corona[8], and protein composition on the nanoparticle surface[9]. However, due to limitations in instrument resolution, the molecular detail of protein-nanoparticle interaction remains poorly understood. Computational modeling, in contrast, provides a useful approach to bridge the gap between experimental observation and the molecular systems of interest. Here we performed both computational and experimental characterisations of protein corona formation between a silver nanoparticle (AgNP) and ubiquitin protein. Silver nanoparticles are widely used in commercial products for their antibacterial and antifungal properties[10], while ubiquitin is ubiquitously expressed in all eukaryotic cells regulating protein distribution and recycling, thereby making AgNP and ubiquitin a representative model system for studying nanoparticle-protein interaction and corona formation.

Two major challenges arise in computational modeling of protein corona. First is the large system size — where an abundance of proteins interacts with nanometer-sized nanoparticles, second is the long timescales associated with protein corona formation. Traditional molecular dynamics approaches can accurately describe the molecular system of nanoparticles and proteins[11, 12], but are not able to reach the relevant time and length scales needed for depicting their interactions till equilibration. To overcome this barrier, we adopted a multiscale modeling approach[13], which coherently blended atomistic and coarse-grained simulations[14, 15]. All-atom simulations were first performed to investigate the possible binding modes between an individual ubiquitin and a AgNP, and the knowledge of AgNP-ubiquitin binding was then incorporated into the construction of a coarse-grained model. With the coarse-grained simulations, we were able to extensively characterise the structure and dynamics of AgNP interacting with multiple ubiquitin molecules (up to 50). The dynamics of both atomistic and coarse-grained models were sampled by discrete molecular dynamics (DMD)[16], an efficient sampling method for underpinning protein dynamics (Supporting Materials).

We first performed atomistic simulations of a molecular system comprised of one ubiquitin molecule and one citrate-coated AgNP (Supporting Materials). The simulations were performed with implicit solvent, and the inter-atomic interactions were modeled by a physical force field adapted from Medusa[17], which include van der Waals, solvation[18], electrostatic, and hydrogen bond potentials. The coarse-grained silver atoms of the AgNP were assigned as hydrophobic with a small fraction being positively charged to account for the nanoparticle surface charges[19]. During simulations, we kept the center of the AgNP static, while allowing the ubiquitin and the citrates to move freely in the simulation box and surface silver atoms mobile on the NP surface.

To evaluate whether ubiquitin could bind to AgNP, we performed DMD simulations near room temperature with a ubiquitin molecule initially positioned away from a citrate-coated AgNP (Fig. 1a). Interestingly, we found that the neutrally-charged ubiquitin did not bind to the hydrophobic surface of AgNP, but instead attracted to the surface charge of the AgNP by replacing the surface-bound citrates (-3e at neutral pH) that were stabilized by electrostatic interactions (Fig. 1a). Although ubiquitin does not have a net charge, it does possess eleven positively-charged and eleven negatively-charged residues out of the 76 total residues[20]. Near the surface of the ubiquitin helix, negatively-charged residues formed a cluster with low electrostatic potentials (Fig. 1b), which allowed stronger binding to the AgNP in simulations than did the negatively-charged citrates. The binding of ubiquitin to AgNP was consistent with our UV-vis absorbance measurement (Fig. 1c), where a redshift from 393 nm (peak wavelength for AgNP absorbance) to 407 nm (peak wavelength for AgNP-ubiquitin absorbance) indicated an increased dielectric constant likely resulting from nanoparticle-protein complex formation. Consistently, our dynamic light scattering measurement showed a hydrodynamic size of 34.5 nm for AgNP-ubiquitin (zeta potential: 12.3 mV), compared to that of 4.8 nm for ubiquitins (zeta potential: 4.6 mV) and 13.6 nm for AgNP (zeta potential: -45.0 mV), further corroborating their effective binding.

To test whether electrostatic interaction was the driving force for AgNP-ubiquitin binding, we artificially enhanced the binding affinity between citrates and AgNP by adding an additional charge to the citrate molecule (Supporting Materials). For both the case of artificially-enhanced electrostatic interactions and the regular (non-enhanced) case, we performed ten independent atomistic DMD simulations with different initial AgNP/ubiquitin configurations. For a higher citrate-AgNP affinity due to enhanced electrostatic interactions, we did not observe any AgNP-ubiquitin binding in all simulations. In the case of regular citrate-AgNP interactions, we observed AgNP-ubiquitin binding for seven out of the ten simulations. The computed distributions of citrates from the AgNP also illustrated that the ability for ubiquitin to displace citrates and bind AgNP depended upon the electrostatic-dominating affinity between the citrates and the AgNP (Fig. 1d). Therefore, the binding of ubiquitin to AgNP was mainly determined by electrostatic interactions.

Based on the ensemble of ubiquitin-bound complex structures derived from seven independent DMD simulations, we computed for each residue the probability of forming contact with the nanoparticle (Supporting Materials). Only a subset of residues had significantly high AgNP contact frequency, $P_{AgNP}$, while the rest of the protein did not interact with the AgNP (Fig. 2a). The histogram of $P_{AgNP}$ featured a bimodal distribution, with one peak close to zero and the other centered around $P_{AgNP} \sim 0.4$ (Fig. 2b). We further determined the AgNP-binding residues (Fig. 2b insert) as those with $P_{AgNP}$ larger than 0.3, the median value separating two peaks in the histogram. These residues were located near the protein helix, coinciding with the cluster of negatively-charged residues (Fig. 1b). Importantly, one of the AgNP-binding residues, Asp18 ($P_{AgNP} = 0.52$), had been experimentally determined to bind gold nanoparticle (AuNP) by nuclear magnetic resonance (NMR) studies[21]. Since AgNP and AuNP are comparable both physically and chemically, we believe that the modes of their binding with ubiquitin are also comparable. This agreement between NMR observations and simulations highlights the predictive power of our computational methods.

In order to observe the formation of AgNP-ubiquitin corona *in silico*, it is necessary to include multiple proteins in simulations, which is beyond the capacity of atomistic simulations. Instead, we used a two-bead-per-residue model[22] to represent ubiquitin and a single atom to model each citrate. The inter- and intra-ubiquitin interactions were modeled by a structure-based potential model[23, 24], which has been extensively used in computational studies of protein folding and protein aggregation[15]. The specific interactions between the AgNP surface charges and ubiquitin residues as well as other non-specific inter-molecule interactions were modeled according to atomistic DMD simulations (Supporting Materials).

We investigated AgNP-ubiquitin corona formation by performing DMD simulations of the coarse-grained system, with multiple ubiquitins (25 molecules) initially positioned randomly with respect to a citrate-coated AgNP. The temperature of the simulation system was kept below the melting temperature of ubiquitin in order to mimic the physiological conditions, where the protein remains folded (Supporting Materials). To avoid potential biases associated with initial conditions, we performed ten independent simulations assuming different initial configurations and velocities. For each simulation we monitored the number of ubiquitins bound to the AgNP, $N_{bound}$, as a function of time. All trajectories in Fig. 3a featured an initial fast binding, which slowed down as time progressed. Interestingly, the average $N_{bound}$ did not follow a typical single-exponential binding kinetics, $\sim 1-exp(-\lambda t)$, which usually features a power-law with the exponent of 1 during initial binding in a log-log plot (Fig. 3b). Instead, the exponent is ~0.23 < 1, corresponding to a stretched-exponential binding kinetics, $\sim 1-exp(-ct^\alpha)$. This unusual behavior might be related to multiple factors that governed AgNP-ubiquitin binding kinetics, including non-specific interactions with other proteins, decreased ubiquitin concentration, depletion of available binding sites for incoming ubiquitins, and competition with citrates. The binding affinity between citrate and AgNP was concentration-dependent, and increased as ubiquitins displaced AgNP-bound citrates to subsequently increase the citrate concentration in solution. All these factors could hinder the binding of ubiquitins to the AgNP surface, leading to the stretched exponential binding kinetics. Therefore, our coarse-grained simulations revealed a rich kinetics for nanoparticle-protein binding, which may need to be considered in future kinetic and mesoscopic modeling of corona formation, such as studies of the Vroman effect of abundant proteins for a nanoparticle entering the bloodstream[25].

The AgNP-ubiquitin complex structure derived from simulations had multiple ubiquitins bound to the surface of one AgNP, forming a single-layer protein corona (Fig. 3c). The majority of AgNP-bound proteins stayed folded under the particular simulation condition (Supporting Materials) and bound to the surface of the AgNP with the protein helix facing the nanoparticle. Only in one of the simulations, one ubiquitin out of the 22 AgNP-bound proteins partially unfolded and the conformation was stabilized by extensive contacts with the hydrophobic surface of the AgNP (Fig. 3c). In addition, we explored the effect of protein concentration on corona formation by performing DMD simulation for a higher ubiquitin/AgNP stoichiometry of 50:1. In these simulations, ubiquitins competed with citrates for binding to the AgNP (Fig. 3d). The final structure featured multiple layers of protein corona, whereas the first layer was dominated by specific binding between ubiquitins and the AgNP, and the outer layers were stabilized by protein-protein interactions (Fig. 3e). The AgNP-ubiquitin complex structures derived from the coarse-grained simulations successfully revealed an atomic picture of the nanoparticle-protein corona.

The ability of nanoparticles to induce protein unfolding (Fig. 3c) could be one of the mechanisms of nanotoxicity. To evaluate the impact of AgNP-binding on ubiquitin conformation, we computed for each protein residue the fraction of native contacts (Q-value[26]) for both the AgNP-bound and unbound ubiquitins (Fig. 4a). A residue with its Q-value close to 1 maintains a native-like structure, while losing its structure if the Q-value is near 0. Both the AgNP-bound and unbound ubiquitins maintained native-like structures with most regions having their Q-values close to 1. Only loop regions between the secondary structures (18-19, 32-35, and 46-53) had relatively low Q-values. The difference in the Q-values for AgNP-bound and unbound ubiquitins suggests that residues in contact with the AgNP were stabilized upon binding (the regions with positive differences coincided with the residues bound to AgNP, Fig. 2a). Two regions, one near the C-terminal of the helix and the other close to residue 46 in a loop, were significantly destabilized upon binding. The destabilization of protein helix due to AgNP-binding is consistent with our circular dichroism (CD) measurement (Supporting Materials), which revealed that the helical content was reduced by 5% for the AgNP-bound ubiquitins compared to the free ubiquitins (Fig. 4b). The increase of β-sheet content could be due to the formation of inter-protein hydrogen bonds between partially unfolded protein regions, since the protein concentration was locally enriched on the AgNP surface.

In summary, both our computer simulations and experiments showed that ubiquitins could readily bind to citrate-coated AgNP. Our multiscale modeling revealed a specific binding between ubiquitins and AgNP, where electrostatic interaction was attributed as the primary driving force. This predicted binding mode is consistent with previous NMR experiments on AuNP-ubiquitin binding. Most notably, our coarse-grained simulations of AgNP-ubiquitin corona formation uncovered an unusual stretched exponential binding kinetics, suggesting complex interactions occurring between AgNP and proteins that may have great relevance to future studies of nanoparticles interacting with biomolecular amphiphiles. At a high stoichiometry, specifically, ubiquitins formed a multi-layer corona surrounding the AgNP. Both our simulations and experiments showed that AgNP-binding destabilized the α-helix while increased the β-sheet content of the ubiquitins. The binding with AgNP altered protein conformation, which may impair recognition of ubiquitin by its binding partners and trigger immune response, act as one of the causes of AgNP-induced cytotoxicity, in addition to ion dissolution that has been accepted as a paradigm for AgNP toxicity[10]. Taken together, our new multiscale modeling has shown a remarkable predictive power for describing the structural and dynamic characteristics of nanoparticle-protein corona, a topic that is poorly understood and yet underlies our interpretation of the transformation and biocompatibility of nanoparticles, and has broad implications in the basic and applied areas of molecular self assembly, nanomedicine, sensing, bioimaging, nanobiophysics, and the health and safety of nanotechnology.

**FIGURES**

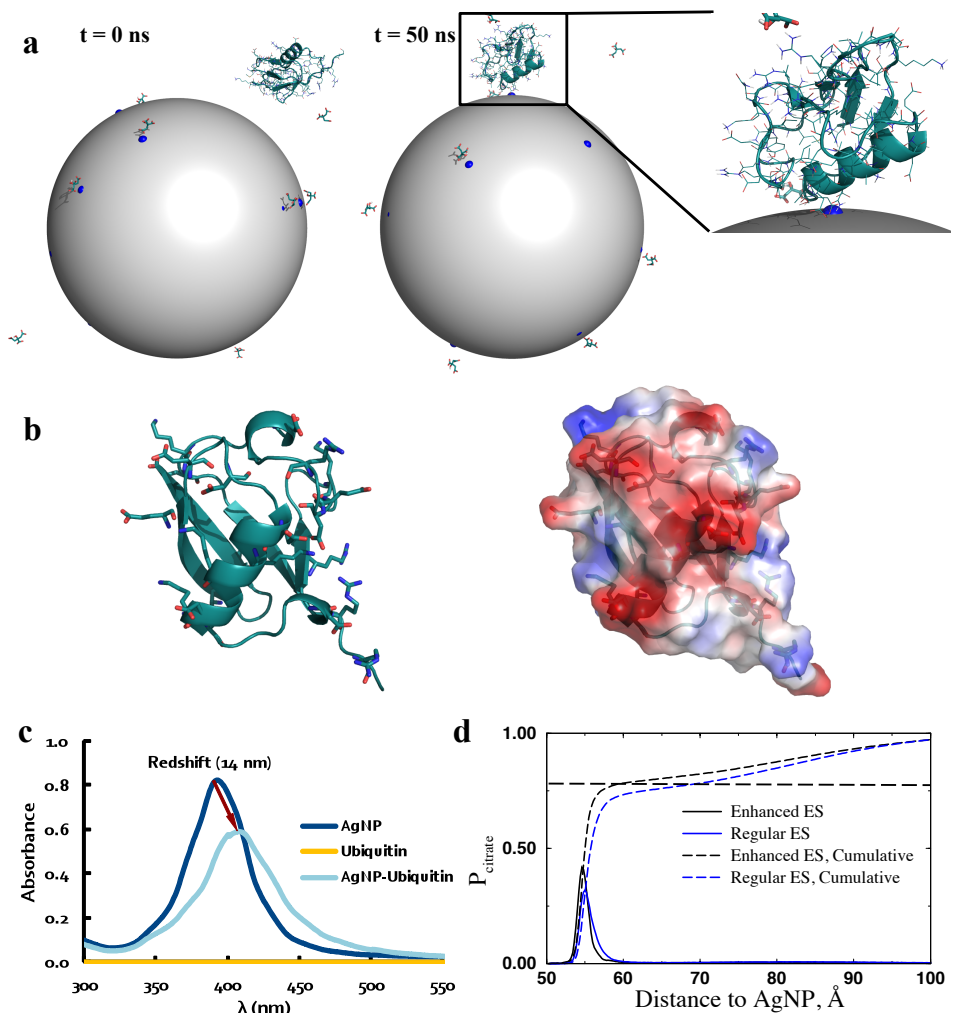

**Figure 1. Interaction between a single ubiquitin and a citrate-coated AgNP.** (a) Initial (t = 0 ns) and final (t = 50 ns) structure of the ubiquitin-citrate-AgNP complex system. The ubiquitin is represented as cartoons, the side chains as lines, and the citrates as sticks. The gray sphere represents the nanoparticle, and the charged atoms on the AgNP surface are shown as blue spheres. Zoom-in view of the final structure indicates the binding between the ubiquitin and a charged AgNP surface atom. (b) The positively (aspartate and glutamate) and negatively (lysine and arginine) charged residues in ubiquitin are shown as sticks (left panel). The surface electrostatic potential (computed using PyMol, www.pymol.org) illustrates the cluster of negatively charged atoms near the protein helix (right panel). (c) The UV-vis absorbance of AgNP, ubiquitin, and AgNP-ubiquitin, featuring a redshift of the absorbance peaks for AgNP-ubiquitin and AgNP alone due to dampened surface plasmon resonance. (d) Distributions of citrates around AgNP (solid lines) derived from the simulations. The electrostatic (ES) interaction between citrate and AgNP was artificially enhanced in one case. The dashed lines correspond to the accumulative probability. The horizontal dashed line corresponds to charge saturation, where the total charge of citrates equal that of the AgNP.

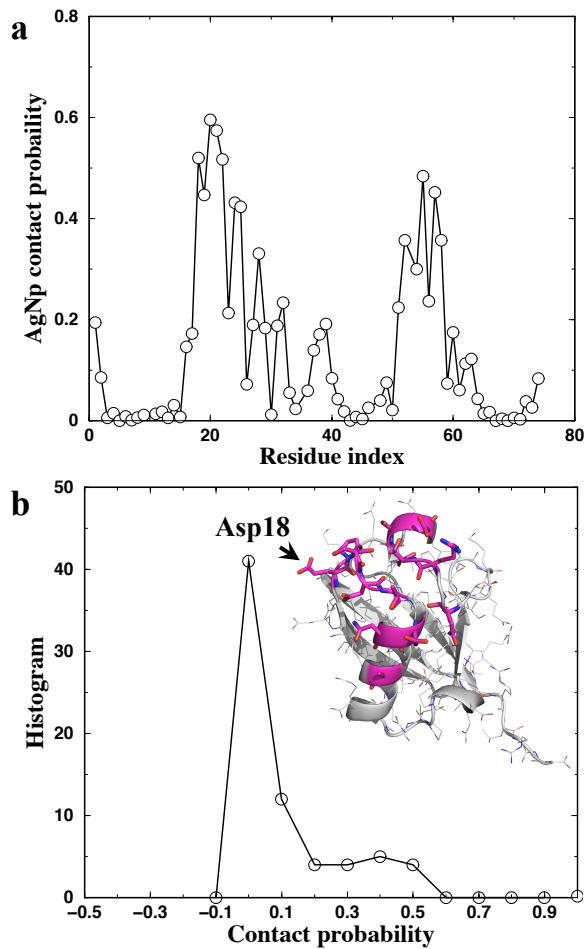

**Figure 2. Specific binding between ubiquitin and AgNP.** (a) The contact probability between AgNP and each ubiquitin residue, computed from independent all-atom DMD simulations (Supporting Materials). (b) The histogram of the AgNP-ubiquitin contact probability displays a bimodal distribution. The ubiquitin residues with high contact frequency (> 0.3; corresponding to the second peak) to the AgNP are shown in sticks (insert). The residue Asp18 was also found to interact with gold nanoparticle[21].

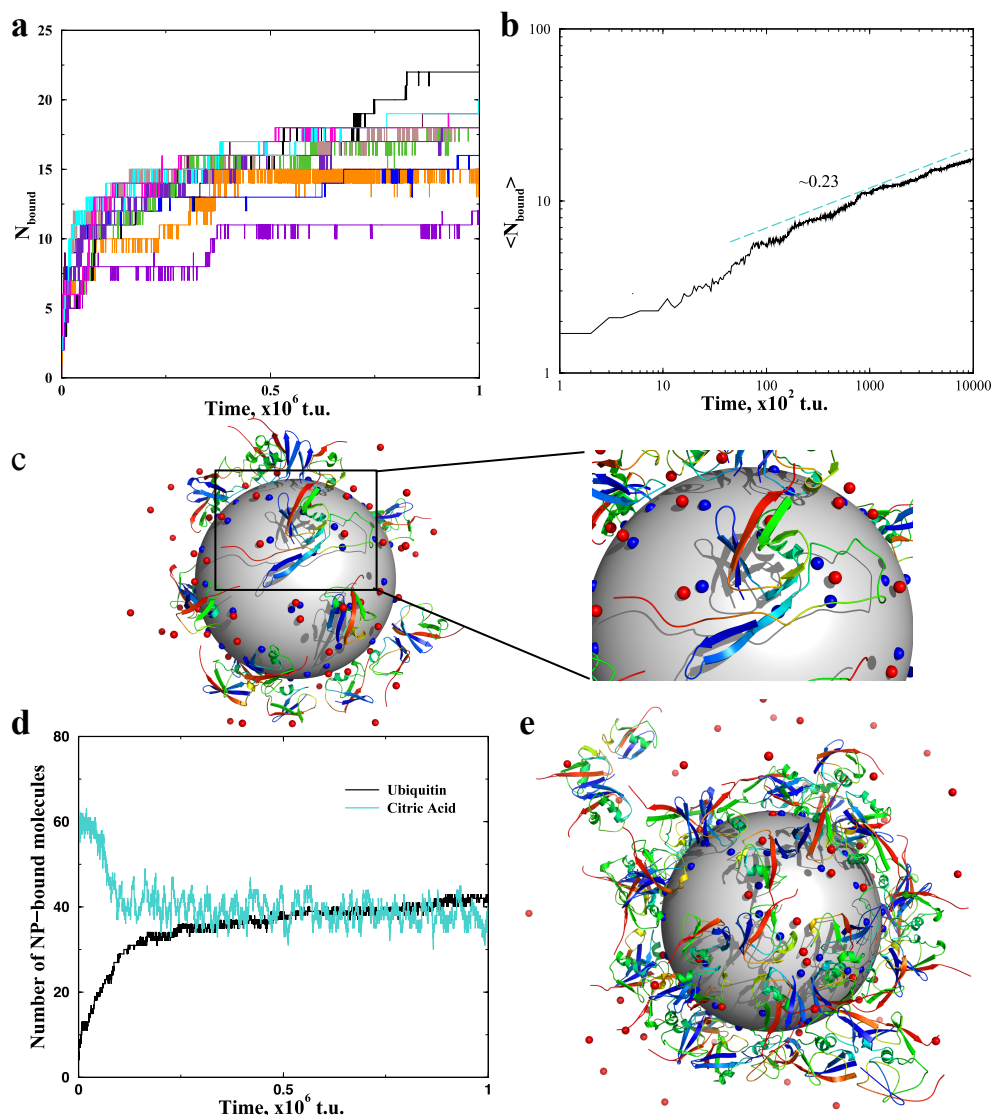

**Figure 3. Ubiquitin-AgNP corona formation.** (a) The number of ubiquitin molecules bound to AgNP, $N_{bound}$, was computed as the function of time (in DMD time unit, t.u., see Supporting Materials) from ten independent simulations (in different colors) of the coarse-grained molecular system. (b) The average number of ubiquitins bound to AgNP, $<N_{bound}>$, features a power-law (approximately linear) in a log-log plot. The exponent is approximately ~0.23. (c) The final structure from one of the simulations (corresponding to the black line with the highest $N_{bound}$ in panel a). The ubiquitins are in cartoon representation. The citrates correspond to the red spheres. The large dark-green sphere denotes the AgNP, and the blue spheres on the surface of the AgNP are the positively charged atoms. One of AgNP-bound ubiquitin is unfolded on the nanoparticle surface (right). In a coarse-grained DMD simulation with a higher stoichiometry of ubiquitin to AgNP (50:1), ubiquitin (black line) competed with citrate (red) to bind AgNP by displacing initially-bound citrates (d). At this high stoichiometry, multi layers of ubiquitins were found to deposit onto the surface of the AgNP (e).

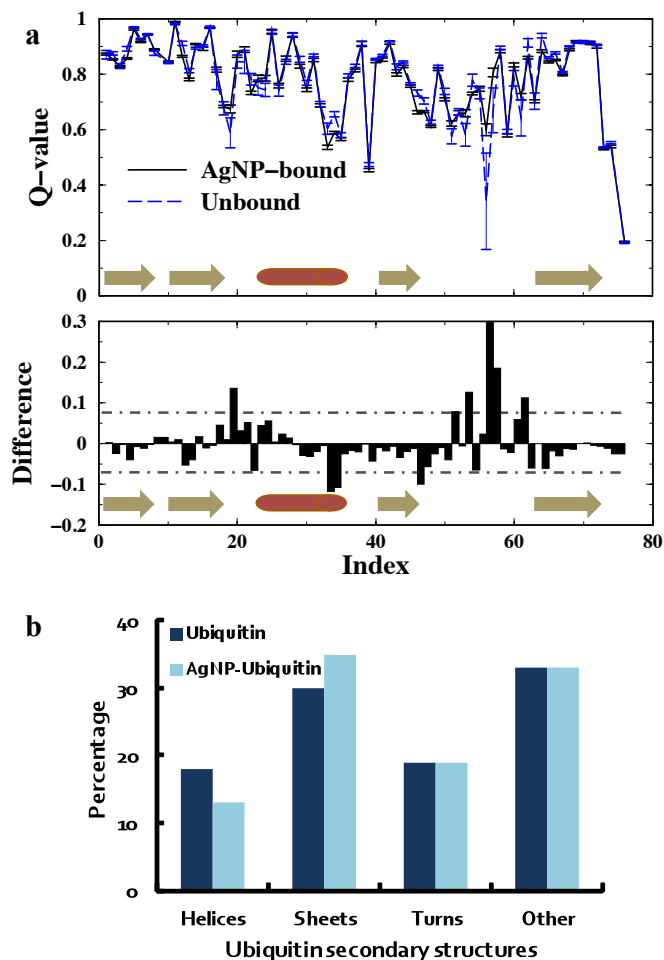

**Figure 4. The structural change of ubiquitin upon AgNP binding.** (a) The fraction of native contacts, Q-value, was computed for each residue for both the AgNP-bound (black) and unbound (blue) ubiquitins (top panel). The error bars were estimated from independent simulations. The yellow arrows indicate the residue segments forming β-strands, and the red rod denotes the residues forming the α-helix. The differences of Q-value were computed between AgNP-bound and unbound (bottom panel) cases. The two dashed lines correspond to deviations with one standard deviation above and below the average. The differences beyond the two lines are statistically significant. (b) The percentage of secondary structures in ubiquitin (dark blue) and in AgNP-ubiquitin (cyan) were probed by CD experiments (Supporting Materials).

**ASSOCIATED CONTENT**
Experimental and Computational Methods are provides as Supporting Materials.

**AUTHOR INFORMATION**
*Corresponding Authors: E-mail: fding@clemson.edu (F.D.) and pcke11@clemson.edu (P.C.K.)

**Notes**
The authors declare no competing financial interest.

## ACKNOWLEDGMENT

The work is supported in part by NSF Grant CBET-1232724 (to P.C.K), an NIEHS grant R01 ES019311 02S1 (to J.M.B. and P.C.K.), and the startup funds from Clemson University (to F.D.). The simulations were performed on the Palmetto high performance cluster, which is managed and maintained by Clemson University CCIT.